\documentclass[aps,prl,showpacs,twocolumn]{revtex4-1}
\usepackage{latexsym}
\usepackage{amssymb}
\usepackage{amsmath}
\usepackage[dvips, pdftex]{graphicx}
\usepackage{tikz}
\newcommand{\ket}[1]{\left| #1 \right>} 
\newcommand{\bra}[1]{\left< #1 \right|} 
\newcommand{\meanv}[1]{\left< #1 \vphantom{#1} \right>} 
 
\newcommand{\abs}[1]{\left| #1 \vphantom{#1} \right|} 
\usepackage{color}
\definecolor{red}{rgb}{1.0,0.0,0.0}
\definecolor{blue}{rgb}{0.0,0.0,1}
\definecolor{green}{rgb}{0.29, 0.33, 0.13}
\begin{document}
\title{Pure dephasing vs. Phonon mediated off-resonant coupling in a quantum-dot-cavity system}
\author{Santiago Echeverri-Arteaga$^{1}$}
\author{Herbert Vinck-Posada$^{1}$}
\affiliation{$^{1}$Departamento de F\'isica, Universidad Nacional de Colombia, 111321, Bogot\'a, Colombia}
\author{Jos\'e M. Villas-B\^oas$^{2}$}
\affiliation{$^{2}$Instituto de F\'isica, Universidade Federal de Uberl\^andia, 38400-902 Uberl\^andia, MG, Brazil}
\author{Edgar A. G\'omez$^{3}$}\email{Corresponding author: eagomez@uniquindio.edu.co}
\affiliation{$^{3}$Programa de F\'isica, Universidad del Quind\'io, 630004, Armenia, Colombia}
\begin{abstract}
Pure dephasing is widely used in the literature to explain experimental observations on quantum dots in cavities. In many cases, its use is not enough and extra terms need to be ``fictitiously'' added to accomplish with the observed data as it is the case of cavity pumping of an unknown source. Here we controvert the validity of the pure dephasing mechanism as a source of decoherence and present a theoretical study based on the phonon-mediated coupling that can explain the emission spectrum and photon auto- and cross-correlation results in recent experiments without the need of any artificial assumptions. We also demonstrated that the phonon-mediated coupling accounts for unexplained features recently reported in measurements of photon auto- and cross-correlation functions. Our work illuminates many of the debates in this field and opens up new possibilities for experimental verification and theoretical predictions.
\\
\end{abstract}
\maketitle
\textit{Introduction.}\textendash
In the context of cavity quantum electrodynamics (cQED), mechanisms such as finite-memory dephasing processes~\cite{Santori:2009}, random charge fluctuations~\cite{Seufert:2000}, spectral diffusion~\cite{Ambrose:1991} and pure dephasing~\cite{Muljarov:2004,Favero:2007} have been recognized as different sources of decoherence in solid-state emitters. Among the physical mechanisms responsible for dephasing, the exciton-phonon interaction in quantum dots (QDs) and those related with phononic process have played an important role and remains a subject of debate. In fact, several research groups have investigated the pure dephasing from both theoretical~\cite{Auffeves:2009,Naesby:2008} and experimental sides~\cite{Borri:2005,Chauvin:2009}, and particularly, the available experimental data reveals that this mechanism increases linearly with temperature, which is a signature of phonon-mediated process~\cite{Banihashemi:2013}. This hypothesis has been mentioned in previous works~\cite{Besombes:2001,Laucht:2009} and is consistent with the fact that the pure dephasing is negligible under conditions of low temperature and excitation power~\cite{Gammon:1996,Cui:2006}. Even though important progress has been made in the field of cQED, some intriguing quantum phenomena within the framework of off-resonant QD-cavity coupling remains not fully understood (and cannot be explained by pure dephasing)~\cite{Hennessy:2007,Tawara:2010,Neuman:2018}. Related experimental measurements on photon correlations have shown a pronounced bunching in the auto-correlation function (under resonance conditions)~\cite{Press:2007} as well as a strong anti-bunching in the cross-correlation function (under off-resonance conditions).~\cite{Kaniber:2008,Ates:2009} Those results also cannot be explained in terms of the pure dephasing, raising many questions about the validity of the pure dephasing mechanism in a regime of low temperatures and it has been conjectured the existence of an unidentified mechanism that possibly is linked to phonon-mediated dephasing~\cite{Kaniber:2008}. Despite the controversy on the role of the pure dephasing, a joint theory-experiment study on dephasing of exciton polaritons has shown that contrary to widely assumed in the literature, pure dephasing could be significant even at very low excitation power densities ($\sim$ 6W/cm$^2$) and temperatures ($\sim $18K)~\cite{Laucht:2009}. In addition, their results showed that the strong coupling regime may appear disguised as a single peak in a QD cavity system and that an unknown source of cavity pumping is needed, opening up a new question.

In this letter we address in details the question of off-resonant coupling in QD cavity system, showing that the phonon-mediated mechanism is able to describes properly the experimental results found in Ref.~\cite{Laucht:2009} without the need of extra pumping terms. We also shown that the pure dephasing mechanism fails to describe recent experimental results on photon auto- and cross-correlation functions, while phonon mediated off-resonant coupling proposed here is more suitable to capture the underlying observed physics and opens up the possibility for new experimental verifications. 

\textit{Theory.}\textendash
We introduce our theoretical model by considering a quantum emitter (QE) coupled to a single cavity mode through the Jaynes--Cummings (JC) Hamiltonian ($\hbar=1$):
$ \hat{H}=\omega_c\hat{a}^\dagger\hat{a}+\omega_x\hat{\sigma}^\dagger\hat{\sigma}+g(\hat{a}^\dagger\hat{\sigma}+\hat{a}\hat{\sigma}^\dagger)$
with $\omega_c$ and $\hat{a}$ ($\hat{a}^{\dagger}$) being the frequency and the annihilation (creation) operator of the single-mode field. Moreover, $\omega_x$ and $\hat{\sigma}=\ket{G}\bra{X}$  is the exciton frequency and corresponding pseudospin operator for the two-level system considering the QD ground state $\ket{G}$ and a single exciton $\ket{X}$ state and $g$ is the coupling constant between cavity mode and exciton. In what follows, we incorporate the influence of the environment through the following master equation
\begin{eqnarray}\label{Master}
\frac{d\hat{\rho}}{dt}&=&-i[\hat{H},\hat{\rho}]+\frac{\Gamma_c}{2}\mathcal{L}_{\hat{a}}(\hat{\rho})+\frac{\Gamma_x}{2}\mathcal{L}_{\hat{\sigma}}(\hat{\rho})+\frac{P_x}{2}\mathcal{L}_{\hat{\sigma}^\dagger}(\hat{\rho})\notag \\
&+&\frac{\gamma_\theta}{2}\mathcal{L}_{\hat{a}\hat{\sigma}^\dagger}(\hat{\rho})+\frac{P_\theta}{2}\mathcal{L}_{\hat{\sigma}\hat{a}^\dagger}(\hat{\rho}),
\end{eqnarray}
where $\mathcal{L}_{\hat{O}}(\hat{\rho})=(2\hat{O}\hat{\rho}\hat{O}^\dagger-\hat{O}^\dagger\hat{O}\hat{\rho}-\hat{\rho}\hat{O}^\dagger\hat{O})$ defines the Lindblad superoperator for an arbitrary operator $\hat{O}$. Notice also that the leakage of cavity photons and the spontaneous emission of the QE are considered through the rates $\Gamma_c$ and $\Gamma_x$, respectively. Additionally, we incorporate in our model the incoherent pumping of the exciton at rate $P_x$ and two effective phononic decay rates $\gamma_{\theta}$ and $P_{\theta}$. In particular, the phononic rates $\gamma_{\theta}$ ($P_{\theta}$) describes the excitation (or de-excitation) of the QE by the annihilation (or creation) of a cavity photon accompanied by the creation (or annihilation) of phonons to compensate the QE-cavity frequency difference. It is worth to mention that these two phononic decay terms have been theoretically proposed by Majumdar and coworkers~\cite{Majumdar:2010} for describing the phonon-mediated off-resonant coupling in cQED systems. 

Our model is significant different from the model presented by Laucht~\textit{et al.}~\cite{Laucht:2009} which includes mechanisms such as incoherent pumping of the cavity mode and the pure dephasing at rates $P_c$ and $\gamma_{x}^\phi$ through the Lindblad terms $(P_c/2)\mathcal{L}_{\hat{a}^\dagger}(\hat{\rho})$, $(\gamma_{x}^\phi/2)\mathcal{L}_{\hat{\sigma}_z}(\hat{\rho})$ into the above master equation (with $\gamma_{\theta}=P_{\theta}=0$). The assumptions of the pure dephasing and incoherent cavity pumping are realistic and well-founded, as it is difficult for the experimentalist to achieve the desired control over aspects such as the proximity of others QE to the cavity mode, as well as processes related to the relaxation of electron-hole pairs and electronic pumping~\cite{LaussyVT:2009}, but there is evidences from both theory and experiments that suggest that pure dephasing does not play an important role for conditions of low excitation power density and very low temperatures~\cite{Nomura:2010}. The cavity pumping is also very controversy~\cite{delValle1:2011}. In fact, those incoherent mechanisms have been proposed in the past as a requirement to reproduce experimental lineshapes~\cite{delValle:2008,Munch:2009,Laucht:2009}. The question that remains is: do they really capture all experimental evidences?

To answer that question, we can compute photoluminescence (PL) spectrum from both models by assuming that most of captured light is coming from the cavity leaking and taking into account the Wiener-Khintchine theorem, which allow us to write the emission spectrum of the coupled system as a Fourier Transform of the two-time correlation function of the operator field $\hat{a}$. More precisely,  $S_a(\omega)=\int_{-\infty}^{\infty} \meanv{\hat{a}^\dagger(\tau)\hat{a}(0)}e^{-i\omega\tau}d\tau$ and for the two-time correlation function should be used the quantum regression theorem~\cite{Perea:2004}. 
\begin{figure}[t!]
\centering
\includegraphics[width=\linewidth]{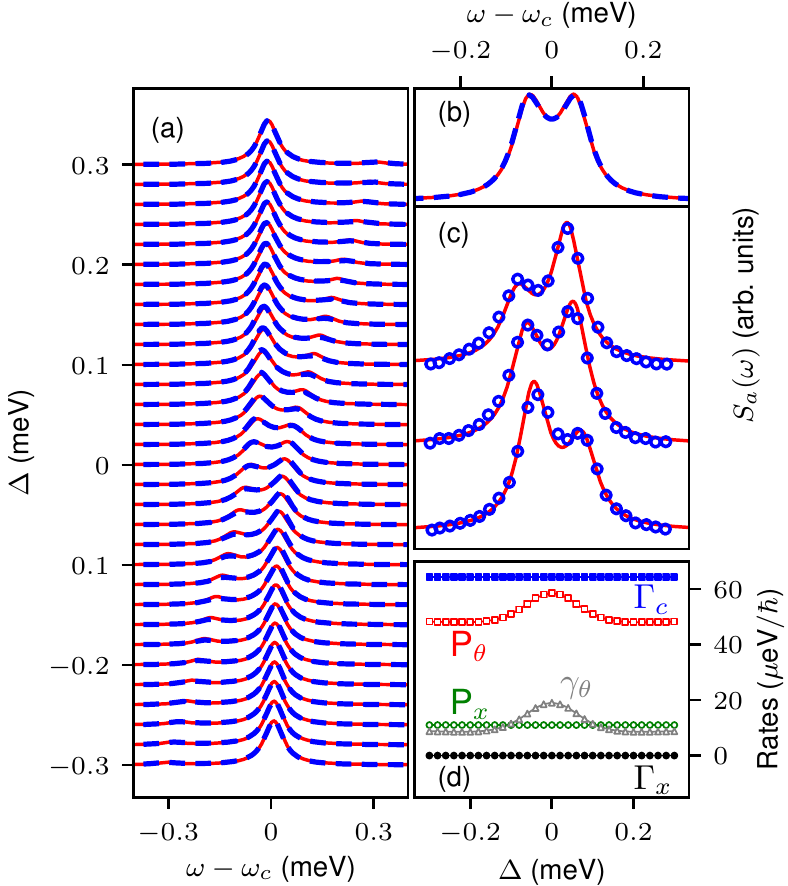}
\caption{(Color online) (a) Emission spectrum for different detunings obtained from model presented in~Ref.~\cite{Laucht:2009} (dashed-blue line) in comparison with our theoretical model (solid-red line) using a fit to find the best parameters. (b) Same as in (a) for the resonant case for better clarity. In (c) we compare the experimental data extracted from Ref.\ \cite{Laucht:2009} (open-blue circles) with numerical results based on our model (solid-red line). Here the detuning are: $\Delta=-50\mu eV$~(top), $\Delta~\sim0\mu eV$~(middle) and $\Delta=+50\mu eV$~(bottom). In (d) we show the dependence of the fitted parameters: $\Gamma_c$ (filled-blue squares), $\Gamma_x$ (filled-black circles), $P_{\theta}$ (open-red squares), $P_{x}$ (open-green circles) and $\gamma_{\theta}$ (solid-gray triangles) on detuning used in (a).}\label{Fig1}
\end{figure}

\textit{Results.}\textendash
In order to examine the underline differences between the two theoretical models, we used the parameters $g=59$ $\mu$eV, $\Gamma_{x}=0.2$ $\mu$eV, $P_{x}=0.5$ $\mu$eV, $\Gamma_{c}=68$ $\mu$eV, $P_{c}=4.5$ $\mu$eV, and $\gamma_{x}^\phi=19.9$ $\mu$eV (which fits experimental data of Ref.~\cite{Laucht:2009}) to generate a set of spectrum for different detunings using the model of Ref.~\cite{Laucht:2009}. Then we perform a numerical fit to it using our new theoretical model. The two set of spectrum are shown in Fig.~\ref{Fig1}(a) and they are in excellent agreement as we can better see in  Fig.~\ref{Fig1}(b), meaning that both models can fit the same experimental PL spectrum data. This is evident in Fig.\ \ref{Fig1}(c), where we compare the experimental data extracted from Ref.~\cite{Laucht:2009} using Engauge software~\cite{Engauge} (open-blue circles) to the PL spectrum obtained from our model (solid-red line). This support the conjecture that the phonon-mediated off-resonant coupling is more suitable to describe the observed effects instead of the pure dephasing, since this mechanism is not expected to be presented at low power and temperature of the fitted experimental data. In Fig.~\ref{Fig1}(d) we show the detuning dependence of the parameters which fits the spectrum shown in Fig.~\ref{Fig1}(a). The results reveals that the parameters $P_x$, $\Gamma_x$ and $\Gamma_c$ remain constant, moreover it corroborates that the spontaneous emission can be neglected as it is usually done in many theoretical studies in the framework of the cQED. The most remarkable result to emerge from these data is that the phononic rates depend on the detuning accordingly to a Gaussian shape and it is consistent with other theoretical models that consider these particular phononic mechanisms at low temperatures $(\sim 4K)$ ~\cite{Roy:2011,Hohenester:2009,Majumdar:2010}. 
\begin{figure}[ht]
\centering
\includegraphics[width=\linewidth]{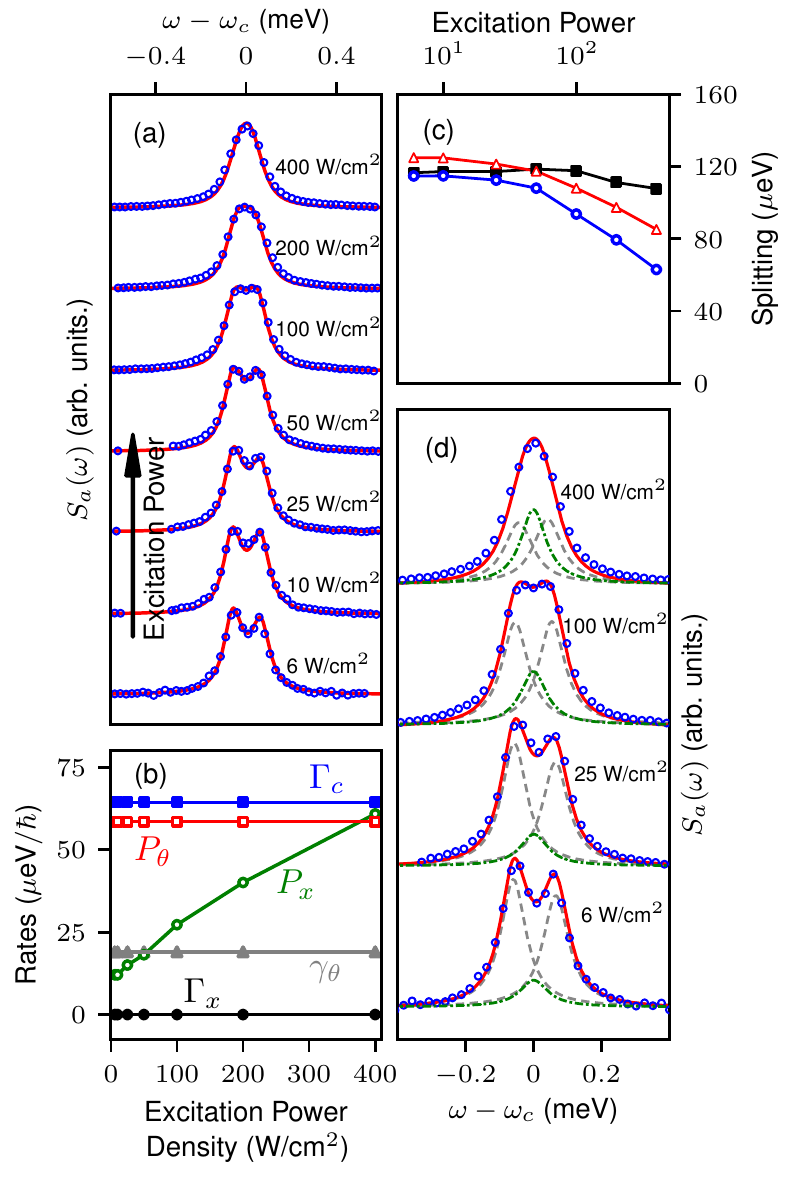}
\caption{(Color online) (a) PL spectrum at resonance as a function of the excitation power. Open-blue circles correspond to experimental data extracted from Ref.~\cite{Laucht:2009}, while the solid-red lines are fits to our theoretical model. In (b) the fitting parameters are displayed as function of the excitation power. The diagram in (c) shows the effective Rabi splitting at resonance as a function of the excitation power. Open-blue circles is difference between the position of the peaks in the PL spectra of Ref.~\cite{Laucht:2009} fitted by two Lorentzian, filled-black squares is the model proposed by Laucht {\it et al.} and open-red triangles is the result of fitting three Lorentzian to our numerically obtained PL spectrum as depicted in (d) for a few situation. In particular, the Rabi doublet is shown as dashed-gray line, whereas the collective state is shown as dot-dashed-green line.}\label{Fig2}
\end{figure}
\\
To further prove that our mode can fit the experimental observation of Ref.~\cite{Laucht:2009}, in Fig.~\ref{Fig2}(a) we fit the extracted PL spectrum for different power densities of that reference using our new model. The parameters found in this fitting is shown in Fig.~\ref{Fig2}(b). As we can see,  $\Gamma_c$ and $\Gamma_x$ stay constant, as we would expect. The only parameter that changes is the exciton pumping rate $P_x$. Surprisingly, we found that the rates $P_\theta$ and $\gamma_\theta$ are independent of the external pump as a clear evidence that the phonon-mediated coupling mechanism must not be attributed to any source of dephasing originated by excitation, in contrast to the pure dephasing rate which increases rapidly with the excitation powers in Ref.~\cite{Laucht:2009}. At the low excitation power limit, the presence of a doublet in the emission spectra is widely considered to be the fingerprint of the strong coupling (SC) regime. However, under certain experimental conditions, depending on relative importance of decoherence in the system, it is possible that the SC regime appears ``in disguise" of a single peak as has been claimed by the authors in Ref.~\cite{Laucht:2009}. 

In Fig.~\ref{Fig2}(c) we display the effective Rabi splitting (ERS) as a function of the excitation powers for different models. Open-blue circles is the difference between the position of the peaks in the experimental spectra of Ref.~\cite{Laucht:2009} fitted by two Lorentzian (traditional approach), while filled black squares are the result of the theoretical model proposed by Laucht~\textit{et al.}. This significant discrepancy in the ERS has been attributed to the broadening in the emission peaks due to the pure dephasing and additional sources of decoherence produced by the cavity pumping. Those source of decoherence was artificially included to fit the spectra and might not represent the true ERS. Therefore, we go further and argue that the change of the vacuum Rabi doublet to a single emission peak can be understood as the result of a dynamical phase transition (DPT) in the system. This phenomena has already been used to explain the unexpected occurrence of an extra emission peak in off-resonant studies and the unusual spectral shifting (cavity-to-exciton attraction) in cQED systems~\cite{Echeverri:2018,Echeverri:2019}. The main consequence of this phenomenon is the coexistence of weak- and strong-coupling regimes, where the vacuum Rabi doublet appears accompanied by a single emission peak at the cavity frequency. This is the result of the formation of a collective state induced by the phonon reservoir. Taking that into account, we fit our theoretical results with three Lorentzian and obtain the ERS as the difference in the peak position of the outer ones  [dashed-gray lines in Fig.~\ref{Fig2}(d)]. The result is show as open-red triangles. In Fig.~\ref{Fig2}(d) dot-dashed-green line represents the Lorentzian of expected collective state.

It is worth mentioning that due to the spectral broadening of the Rabi doublet, it imposes a limit in the spatial resolution and therefore the experimental observation of the collective state becomes a challenge for the experimentalist. In fact, several experiments related to QE-cavity system could not observe such phenomenology~\cite{Laucht:2009,Nomura:2010,Muller:2015,Sek:2006}. 
From theory we expect that for large ratio of $g/\abs{\Gamma_c-\Gamma_x}$ and a very low cooperativity factor $C=g^2/\Gamma_c\Gamma_x$, it is possible to find the appropriated experimental conditions where a spectral triplet will appears in the PL spectrum. Taking into account the current experimental scenario only few researchers have reached the above mentioned conditions and consequently the spectral triplet has been observed~\cite{Hennessy:2007,Ota:2009}. 

Another kind of experiments that might hint that pure dephasing is not the correct way to treat the decohence in this system is the auto and cros-correlation measurements as found in Refs.~\cite{Laucht:2011,Kiraz:2002,Calic:2011,Winger:2009,Ates:2009}. They have found a strong photon anti-bunching in both kind of measurements. The origin of the strong photon anti-bunching in the cros-correlation function was the subject of many debates and phonon-mediated process spears as the most probable candidate. In order to quantify theoretically the photon correlations, we consider the auto- and cross-correlation function in the steady state defined by $g^{(2)}(\tau)=\meanv{\hat{a}^{\dagger}(0)\hat{a}^{\dagger}(\tau)\hat{a}(\tau)\hat{a}(0)}/\meanv{\hat{a}^\dagger\hat{a}}^2$ and $g_X^{(2)}(\tau)=\meanv{\hat{a}^{\dagger}(0)\hat{\sigma}^{\dagger}(\tau)\hat{\sigma}(\tau)\hat{a}(0)}/\meanv{\hat{a}^\dagger\hat{a}}\meanv{\hat{\sigma}^\dagger\hat{\sigma}}$, respectively. 

\begin{figure} 
\centering
\includegraphics[width=\linewidth]{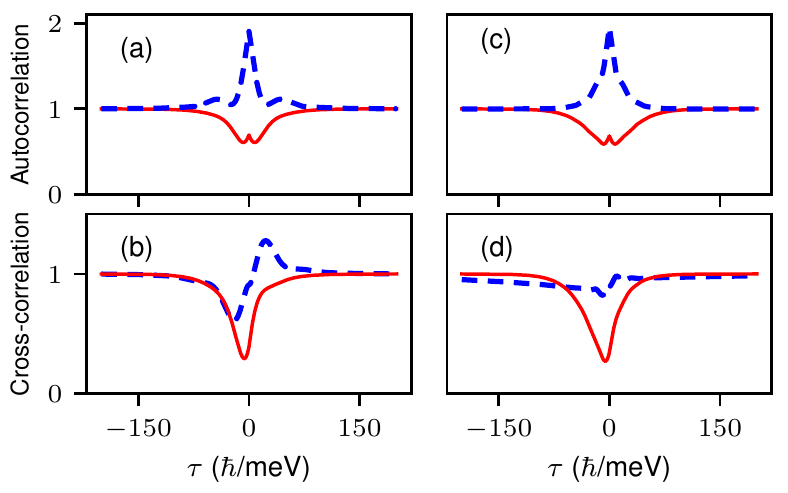}
\caption{(a) and (b) shows the  auto- and cross-correlation function for $\Delta$=0, while (c) and (d) show the same for $\Delta=0.3meV$.  Solid-red line is our model and dashed-blue line is for model used in Ref.~\cite{Laucht:2009}. Here we used the parameters of Fig.~\ref{Fig1}.
}\label{Fig3}
\end{figure}
Fig.~\ref{Fig3}(a)-(b) shows the numerically calculated auto- and cross- correlation function at the resonance condition. The numerical results based on our theoretical model are shown with solid-red line, whereas the results using the model presented by Laucht~\textit{et al.} are shown with dashed-blue line. It is interesting to note that our model is in qualitatively agreement with experimental results previously reported by other authors~\cite{Ates:2009,Kaniber:2008,Laucht:2011}. It can be seen a strong anti-bunching in the photon correlations together with some particular features such as a small peak around $\tau=0$ in the auto-correlation function, which is a signature of ``an unknown intermediate coupling regime''~\cite{Dubin:2010}, and a pronounced asymmetry in the cross-correlation function that has not been explained theoretically so far~\cite{Hennessy:2007,Kaniber:2008,Kiraz:2002,Calic:2011}. The out of resonance ($\Delta=0.3meV$) results are shown in Fig.~\ref{Fig3}(c)-(d). It is interesting to notice that the features in the auto- and cross- correlation functions described above holds for larger detunings, in agreement with experimental observations, and contrary to the established in the literature, this unexpected behavior can be understood from a Markovian perspective~\cite{Ates:2009,Kaniber:2008,Kiraz:2002,Calic:2011,Winger:2009}. The result based on pure dephasing model (dashed-blue line) fails completely in describing what is observed experimentally.

\textit{Conclusions.}\textendash
The findings of this study suggest that contrary to a claim in the literature, the phonon-mediated off-resonant coupling could be the suitable mechanism for describing decoherence at very low temperatures instead of the pure dephasing mechanism. Our theoretical calculations based on phononic mechanisms agree well with the current experimental observations. More precisely, we found that our results are in qualitative agreement with the available experimental data of the PL spectrum of a QE-cavity system, as well as they are consistent with some unexpected features found in experimental measurements of the auto- and cross- correlations functions. Although our theoretical model confirms that even when the QE-cavity system has a single emission peak in the PL spectrum, it is possible to identify the ERS as a signature of the strong coupling regime. Moreover, it  
exhibits an apparent significant reduction as a consequence of the emergence of a collective state at the cavity frequency. Our findings support the idea that there is an intermediate coupling regime which has been theoretically predicted as a DPT in the system~\cite{Echeverri:2018}, and therefore the experimentalists shall be encouraged to search for more evidences of the novel coupling regime in cQED systems.

\textit{Acknowledgments.}\textendash
S.E.-A. and H.V.-P. gratefully acknowledge funding by COLCIENCIAS projects  ``Emisi\'on en sistemas de Qubits Superconductores acoplados a la radiaci\'on'', c\'odigo~110171249692, CT~293-2016, HERMES 31361, ``Control din\'amico de la emisi\'on en sistemas de qubits acoplados con cavidades no-estacionarias'', c\'odigo~201010028998, HERMES~41611 and ``Interacci\' on radiaci\'on-materia mediada por fonones en la electrodin\'amica cu\'antica de cavidades'', c\'odigo~201010028651, HERMES~42134. S.E.-A. also acknowledges support from the  ``Beca de Doctorados Nacionales'' de COLCIENCIAS convocatoria~727. E.A.G acknowledges financial support from Vicerrector\'ia de Investigaciones of the Universidad del Quind\'io through the project No.~919. JMVB acknowledge the financial support of CNPq (Grant No. 462123/2014-6 and 308929/2015-2)

\end{document}